\documentclass[10pt,apjl,tighten]{emulateapj}

\usepackage{graphicx}
\usepackage{tabulary}
\usepackage{epsfig}
\usepackage{amssymb}
\usepackage{multirow}
\usepackage{appendix}
\usepackage{natbib}
\usepackage{wasysym}
\usepackage[pdftitle={Rotational Synchronization May Enhance Habitability for
  Circumbinary Planets: Kepler Binary Case Studies}, colorlinks = true, breaklinks = true,
  citecolor = blue, linkcolor = blue, urlcolor = blue, pdfauthor =
  {Jorge I. Zuluaga}]{hyperref}
\newcommand{\beq}[1]{\begin{equation}\label{#1}}
\newcommand{\eeq}{\end{equation}}

\newcommand{\Mdot}{\mbox{\.{M}}}
\newcommand{\Msun}{\mbox{$M_{\odot}$}}

\newcommand{\Pbin}{P_{\rm bin}}
\newcommand{\Ms}{M_\star}
\newcommand{\Prot}[1]{P_{\rm rot,#1}}
\newcommand{\sub}[1]{_{\rm #1}}

\renewcommand{\Sun}{\odot}
\renewcommand{\sun}{\odot}

\shorttitle{Rotational Synchronization May Enhance Habitability for
  Circumbinary Planets: Kepler Binary Case Studies}

\shortauthors{Paul A. Mason, Jorge Zuluaga, Joni Clark \& Pablo A. Cuartas}

\begin{document}

\title{Rotational Synchronization May Enhance Habitability for
  Circumbinary Planets: Kepler Binary Case Studies}

\author{Paul A. Mason\altaffilmark{1}}

\affil{Department of Physics, University of Texas at El Paso, El Paso,
  TX 79968, USA}

\affil{Department of Mathematics and Physical Sciences, New Mexico
  State University - DACC, Las Cruces, NM, 88003, USA}

\author{Jorge I. Zuluaga\altaffilmark{2}}

\affil{FACom - Instituto de F\'{\i}sica - FCEN, Universidad de
  Antioquia, Calle 70 No. 52-21, Medell\'{\i}n, Colombia}

\author{Joni M. Clark\altaffilmark{3}}

\affil{Department of Mathematics and Physical Sciences, New Mexico
  State University - DACC, Las Cruces, NM, 88003, USA}

\and

\author{Pablo A. Cuartas-Restrepo\altaffilmark{4}}

\affil{FACom - Instituto de F\'{\i}sica - FCEN, Universidad de
  Antioquia, Calle 70 No. 52-21, Medell\'{\i}n, Colombia}

\altaffiltext{1}{pmason@nmsu.edu}
\altaffiltext{2}{jzuluaga@fisica.udea.edu.co}
\altaffiltext{3}{jonic@nmsu.edu}
\altaffiltext{4}{p.cuartas@fisica.udea.edu.co}

\medskip

\begin{abstract}
We report a mechanism capable of reducing (or increasing) stellar
activity in binary stars, thereby potentially enhancing (or
destroying) circumbinary habitability.  In single stars, stellar
aggression towards planetary atmospheres causes mass-loss, which is
especially detrimental for late-type stars, because habitable zones
are very close and activity is long lasting. In binaries, tidal
rotational breaking reduces magnetic activity, thus reducing harmful
levels of XUV radiation and stellar mass-loss that are able to erode
planetary atmospheres. We study this mechanism for all confirmed
circumbinary (p-type) planets.  We find that main sequence twins
provide minimal flux variation and in some cases improved
environments, if the stars rotationally synchronize within the first
Gyr. Solar-like twins, like Kepler 34 and Kepler 35, provide low
habitable zone XUV fluxes and stellar wind pressures.  These wide,
moist, habitable zones may potentially support multiple habitable
planets. Solar-type stars with lower mass companions, like Kepler 47,
allow for protected planets over a wide range of secondary masses and
binary periods. Kepler 38 and related binaries are marginal
cases. Kepler 64 and analogues have dramatically reduced stellar
aggression due to synchronization of the primary, but are limited by
the short lifetime. Kepler 16 appears to be inhospitable to planets
due to extreme XUV flux. These results have important implications for
estimates of the number of stellar systems containing habitable
planets in the Galaxy and allow for the selection of binaries suitable
for follow-up searches for habitable planets.
\end{abstract}


\section{Introduction}
\label{sec:introduction}

Seven confirmed planets have been found orbiting six binary stars 
\citep{Doyle11, Orosz12a, Orosz12b, Welsh12, Schwamb13}, igniting
interest in the possibility of terrestrial planets in circumbinary
radiative habitable zones (RHZs) \citep{Mason12, Kane13, Clark13}.
Planetary atmosphere erosion by stellar winds and intense XUV fluxes
must also be considered when assessing circumbinary
habitability. These erosive processes could obliterate planetary
atmospheres or produce desiccation, akin to Venus \citep{Zuluaga13}.
In this letter, we present a mechanism by which stellar activity in
binaries can be reduced (increased) due to tidal breaking of stellar
components, potentially enhancing (restricting) the protection of
planetary atmospheres against these stellar aggression factors.

Relations between age, rotation rate, and magnetic activity of single
stars have been established both theoretically and observationally
\citep{Basri87,Wood05}. Rapidly rotating stars are luminous XUV
sources and undergo significant mass-loss \citep{Wood05}, posing high
risks to weakly magnetized planets. These relationships have been used
to evaluate the evolution of stellar aggression and its role in
terrestrial planet habitability \citep{Griebmeier07, Zuluaga13}. When
these relationships are applied to binaries, we find that early tidal
spin-down of one or both stars produces an effective stellar
rotational aging. Thus reducing stellar aggression, abating mass-loss
from planetary atmospheres, and potentially promoting habitability. In
other cases the stellar aging effect increases activity and reduces the
chances for habitability.

To explore this effect, we extend single star RHZ limits
\citep{Kasting93, Kopparapu13} to planets in circumbinary orbits
(Section \ref{sec:circumbinary}) and investigate the magnitude of
stellar aggression towards these planets (Section
\ref{sec:planet-binary-interaction}). An ensemble of main sequence
binaries with components from 0.2$M_{\Sun}$ to 1.5$M_{\Sun}$, and a
range of binary periods and eccentricities are modeled (Section
\ref{sec:results}). Synchronization times are computed for stars in
both twin and disparate binaries, including six Kepler binaries with
known circumbinary planets (Section \ref{sec:results}).

\section{Circumbinary Habitability}
\label{sec:circumbinary}

Circumbinary habitability is a five dimensional parameter problem,
even for circular planetary orbits.  These are the
mass of the primary and secondary, $M_1$ and $M_2$, binary
eccentricity and period, $e$ and $P_{bin}$, and planetary semi-major
axis $a$. Five dimensions are reduced to four by considering only
orbits in the middle of the circumbinary habitable zone.

Two cross-sections of the remaining four variable problem are
examined: (1) twins, i.e. $M_1 = M_2$ and (2) binaries with a solar
mass primary and a lower mass companion. Examination of these
cross-sections and specific case studies provided by the six
Kepler binaries elucidate circumstances for which enhanced, or reduced,
habitability induced by the tidal breaking mechanism operates.


\subsection{Orbital Stability}
\label{subsec:orbital-stability}

Orbital stability is a prerequisite for habitability.  To be stable,
the semi-major axis of a circumbinary planet must be larger than a
critical value $a_{c}$.  For a large range of binary semi-major axes
$a_{\rm bin}$ and eccentricities $e$, $a_c$ is calculated from
numerical fits, such as that provided by Eq. (3) of \citet{Holman99}.

Orbital stability within the RHZ is of greatest concern for similar
mass binaries with large separations. For twin binaries in circular
orbits, i.e. $\mu=1/2$ and $e=0$, the stability criterion simplifies
to $a_{c} \sim 2.4\,a_{\rm bin}$.  In this case, if $a_{\rm
  bin}>l\sub{out}/2.4$, where $l\sub{out}$ is the outer edge of the
RHZ, planets throughout the habitable zone have unstable orbits,
rendering the binary uninhabitable.


\subsection{Circumbinary Radiative Habitable Zone}
\label{subsec:hz}

RHZ limits are found by calculating fluxes which allow
the existence of liquid water.  We follow the
results of \citet{Kopparapu13} that, for single stars, provide
limiting fluxes $S\sub{eff}$ (in units of present Earth solar flux)
as a function of stellar effective temperature $T_{*}=T_{eff}-5780 K$:

\begin{equation}
\label{eq:Seff}
S\sub{eff}=S_{{\rm eff}\odot}+aT_{*}+bT^{2}_{*}+cT^{3}_{*}+dT^{4}_{*}
\end{equation}

Here $S\sub{eff\odot}$, $a$, $b$, $c$ and $d$ are constants which
depend on the physical criterion defining a given limit and are
tabulated in Table 3 of \citet{Kopparapu13}.  The most generous limits
are obtained by assuming that Venus had surface liquid water until a
few Gyr ago (the recent Venus criterion) and that Mars was also
habitable at the beginning of solar system evolution (the early
Mars criterion).

Limits of the RHZ, either around single stars or binaries are
defined as the distance $d$ where the averaged flux $\langle
S(d)\rangle$ is equal to the critical flux:

\beq{eq:HZ}
\langle S(d)\rangle = S\sub{eff}
\eeq

For single stars, circular orbits, and fast rotating planets, $\langle
S(d)\rangle=L_{*}/d^2$, where $L_{*}$ is the stellar luminosity in
solar units and $d$ is in AU.

In order to apply Eq. (\ref{eq:Seff}) to binaries we first
verify that an effective temperature can be associated with the
combined stellar flux.  For twins, the result is trivial,
but for disparate binaries, a more complex situation arises.  We have
verified numerically that the peak flux in disparate systems remains
close to that of the primary. So conservatively, we assume that
the effective temperature is that of the most
luminous star, since it has the dominant effect on the RHZ.

The flux at a distance $d$ from the center of mass (CM), varies with
the binary phase angle $\theta$ as:

\beq{eq:BinaryFlux}
S\sub{bin}(d,\theta)=\frac{L_1}{R_1^2(d,\theta)}+\frac{L_2}{R_2^2(d,\theta)}
\eeq

where $R_1$ and $R_2$ are planetary distances to the primary and
secondary components.  Assuming that the planetary orbit is in the
same plane as the binary,

\begin{eqnarray*}
R_1^2(d,\theta) & = & (d+r_1\sin\theta)^2+r_1^2\cos^2\theta\\
R_2^2(d,\theta) & = & (d-r_2\sin\theta)^2+r_2^2\cos^2\theta
\end{eqnarray*}

Here $r_2=a\sub{bin}/(1+q)$ and $r_1=q r_2$ are the average star-CM
distances and $q = M_2 /M_1$ is the binary mass ratio.  To calculate
RHZ limits we compare the combined flux (Eq. \ref{eq:BinaryFlux})
averaged over the binary orbit, with the effective flux for a given
RHZ edge:

\beq{eq:HZbin}
\frac{1}{2\pi}\int_0^{2\pi}\left(\frac{L_1}{R_1^2(d,\theta)}+
\frac{L_2}{R_2^2(d,\theta)}\right)d\theta=S_{eff}
\eeq

\begin{figure}
\epsscale{1.15}
\plotone{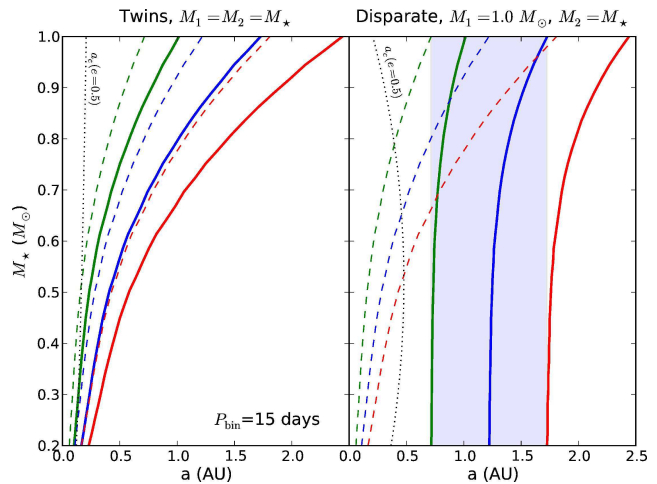}
\caption{Habitable zones for single stars (dashed lines) and for
  binaries (solid lines) are shown for twins (left) and solar
  primaries (right).  The vertical axis shows the mass in the twin
  cases and the secondary mass for the solar primary cases.  The
  binary period is 15 days. The inner RHZ edge is the recent Venus
  limit (green) and the outer edge is the early Mars limit (red).  The
  average of these extremes is shown in blue. The solar system is
  shown as the shaded region, on the right, corresponding to the limit
  of the binary RHZ as the secondary mass approaches 0.0 $M_{\Sun}$.
  Critical distances for orbital stability assuming an $e=0.5$ binary
  orbit are also shown for reference.}
\label{fig:binaryHZ}
\end{figure}

The result of solving Equation (\ref{eq:HZbin}) in the case of twins
and disparate binaries is presented in Figure \ref{fig:binaryHZ}.

\bigskip

Orbital stability and proper insolation do not guarantee
the most important condition for habitability: the presence of a dense
and wet atmosphere.  Although important intrinsic factors
are involved in the formation and preservation of planetary
atmospheres, planet-star interactions play key roles in the fate of
gaseous planetary envelopes, especially during the early, active,
phases of stellar evolution.  In the next section, we model
several aspects of planet-star interactions and apply them to the
survival of circumbinary planet atmospheres.

\section{Planet-stars interaction}
\label{sec:planet-binary-interaction}

Winds from low-mass stars ($M_\star\lesssim 1M_\odot$) play a
central role in planetary atmosphere retention.  
Planets could lose their atmospheres or be desiccated on a
time-scale much less than that required for the evolution of complex
life \citep{Zendejas10,Lammer12}.  Even if planets have magnetic
fields comparable to Earth, but are located in the RHZ of K-M stars,
they could be subject to intense XUV irradiation \citep{Segura10}. The
resulting water loss is important, especially in the case of developing
Nitrogen-rich atmospheres during early phases of planetary evolution
\citep{Lammer09}.  This is not a second-order effect, but can become
the dominant factor determining habitability of terrestrial planets.

\subsection{Stellar activity and rotational aging} 
\label{subsec:binary-sw}

A rigorous treatment of binary stellar winds is challenging
\citep{Siscoe1974}.  The problem has been extensively studied in the
case of early type binaries \citep{Stevens92}, but less attention has
been paid to the case of low-mass stars in binaries.  Here, we assume a
simplified non-interacting model for the combined stellar wind.

Since for binary periods of $P_{bin}\sim 10-20$ days, orbital
velocities ($v\sim$ 80-100 km/s) are much lower than wind velocities
as measured near the stellar surface ($v_{\rm sw}>$ 3000 km/s), we
neglect orbital motion effects in calculating stellar wind properties.

For twins, we assume a wind source with a coronal temperature and wind
velocity profile equal to that of a single star.  Mass-loss rates are
assumed to be double of that calculated for single stars of the same
type. For disparate binaries, we simply sum the stellar wind pressure
from each component.

Stellar wind properties are calculated using Parker's model
\citep{Parker58}.  It has been shown \citep{Zuluaga13} that planetary
magnetospheric properties computed with this model are only $\sim
10\%$ different than those obtained with more realistic models.

The stellar wind's average particle velocity, $v_{\rm sw}(d)$, at a
distance $d$ from the host star is obtained by solving Parker's
equation (Equation (35) in \citealt{Zuluaga13}).  The density profile,
$n_{\rm sw}(d)$ is obtained by equating particle velocity and
mass-loss rate $\dot{\Ms}$:

\begin{equation}
n_{\rm sw}=\frac{\dot{\Ms}}{4\pi d^2 v_{\rm sw} m}
\end{equation}

A procedure to estimate Parker model parameters as a function of
stellar age for single stars was devised by \citet{Griebmeier07}.
It relies on empirical relationships between age, X-ray
flux, and rotation of single stars \citep{Wood05}. According to these
relations the product of mass-loss rate and wind velocity $\Mdot v\sub{sw}$
scale with stellar rotation period $P\sub{rot}$ following:

\beq{eq:Protscaling}
\Mdot v\sub{sw}\propto P\sub{rot}^{-3.3}
\eeq

We adapt these results to binaries by introducing the so called
``rotational age'', $\tau_{\rm rot}$ defined as the
equivalent stellar age at which rotational period of a single star
$\Prot{single}$ will be equal to rotational period of a star in the
binary $P\sub{rot,bin}$:

\begin{equation}
\label{eq:rotage}
\Prot{single}(\tau_{\rm rot})=P\sub{rot,bin}
\end{equation}

For non-negligible binary eccentricity, stars
eventually reach a pseudo-synchronous state with
$P\sub{rot,bin}=\Pbin/n\sub{sync}$ (see Section
\ref{subsec:tidal-interaction}) where $n\sub{sync}$ is a real number
depending on eccentricity\citep{Hut81}.

Applying these isolated star relationships to binaries is
reasonable since physical mechanisms connecting rotation, age, and
activity would not be different for stars in binaries with separations of
tens to hundreds of stellar radii, i.e. for $P\sub{bin}>5$ days.

\begin{figure}
\epsscale{1.15}
\plotone{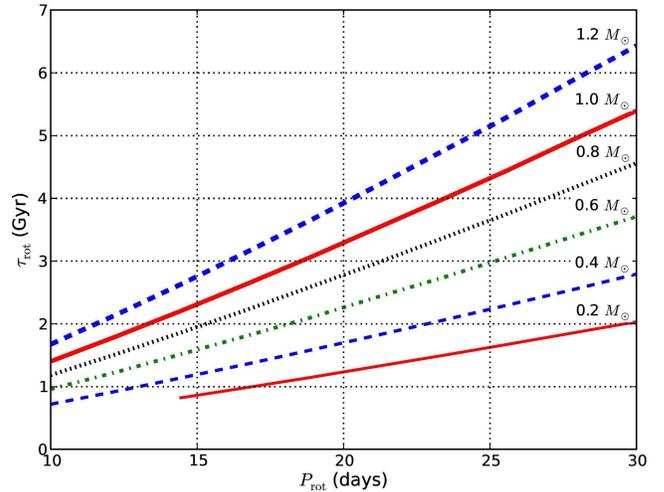}
\caption{Rotational ages for single stars as a function of rotational
  period.  We assume that a similar relationship applies to stars in
  binaries where the rotational period is replaced by a multiple of the
  binary period.}
\label{fig:rotational-aging}
\end{figure}

Rotational ages for stars in binaries  are depicted in 
Figure \ref{fig:rotational-aging}.  We see
that F, G and late K stars ($M_\star>0.8$) in binaries with orbital
periods $P\sub{bin,rot}<20$ days, could experience premature aging if
they are quickly tidally locked.  They would appear as
old as single stars with $\tau>3$ Gyr, in terms of magnetic and
stellar activity.  Premature aging might be an advantage for 
circumbinary terrestrial planet atmospheres.  On the other hand,
lower mass stars in binaries with similar periods exhibit a
forever-young effect, i.e. components freeze at rotational ages less
than approximately 2 Gyr, thereby reducing habitability.

Stellar XUV luminosity depends on chromospheric and coronal activity,
which in turn depends on rotation.  Since rotation of single MS
stars slows down with age, XUV luminosity should also decrease with
time \citep{Garces11}.  The rotational aging mechanism will reduce 
XUV luminosities and their potential harmful effects
on planetary atmospheres.

Despite large uncertainties in measured XUV stellar emission
\citep{Pizzolato03}, several authors have developed simple empirical
laws expressing XUV luminosities, or its proxy, X-ray luminosity, as a
function of age.  To be conservative, we use the law by
\citealt{Garces11} providing X-ray luminosity of GKM-types as a power-law
of stellar age:

\beq{eq:LXFunc}
L_X=\left\{
\begin{array}{ll}
6.3\times 10^{-4} L_\star & \rm{if}\;\tau<\tau_i \\
1.89\times 10^{28}\;\tau^{-1.55} & \rm{otherwise}
\end{array}
\right.
\eeq

where $L_\star$ is the bolometric luminosity and $\tau_i$ is the so
called saturation time scaling with $L_\star$ according to:

\beq{eq:ti}
\tau_i=0.06\,\rm{Gyr}\;\left(\frac{L_\star}{L_\odot}\right)^{-0.65}
\eeq

For high X-ray luminosities we approximate $L\sub{XUV}\approx
L\sub{X}$ \citep{Guinan09}. Using this model, we verify that XUV
Present Earth Level (PEL) is 0.88 erg cm$^{-2}$ s$^{-1}$, in agreement
with the observed value \citep{Judge03}.  We also predict that at
$\tau_\odot\approx 1$ Gyr the Earth XUV flux was $F\sub{XUV}=8$ PEL in
agreement with previous estimates (see e.g. \citealt{Kulikov06}).

\subsection{Binary tidal interaction}
\label{subsec:tidal-interaction}

Benefits are maximized if tidal locking occurs,
at least for the primary component, before the rise
of the secondary planetary atmosphere.  For solar system terrestrial
planets, this time is estimated as $\tau\sub{atm}\sim
0.3-0.8\,{\rm Gyr}$ \citep{Hart78,Hunten93}.  Therefore, in order to
evaluate if circumbinary planets benefit from rotational aging we
estimate synchronization times.

For a target star with initial rotational and orbital angular
velocities $\Omega=2\pi/{P\sub{rot}}$ and $\omega=2\pi/\Pbin$, subject
to secondary star tides placed in an orbit with
eccentricity $e$, synchronization time $t_{sync}$ is calculated
following \citet{Zahn08},

\begin{eqnarray}
\label{eq:tsync}
\frac{1}{t_{sync}}&=&\frac{1}{t_{diss}}
\frac{f_2(e^2)}{(1-e^2)^6}\times\\\nonumber
& & \times\left[1-\frac{(1-e^2)^{3/2}f_5(e^2)}{f_2(e^2)}\frac{\omega}{\Omega}\right]\times\\\nonumber
& & \times\left(\frac{M\sub{field}}{M\sub{targ}}\right)^{2}\frac{M\sub{targ}R\sub{targ}^{2}}{I}\left(\frac{R}{a}^{6}\right)
\end{eqnarray}

where:

\begin{eqnarray*}
f_2(e^2) & = & 1+\frac{15}{2}e^2+\frac{45}{8}e^4+\frac{5}{16}e^6\\
f_5(e^2) & = & 1+3e^2+\frac{3}{8}e^4
\end{eqnarray*}

Here $I$ is the moment of inertia of the target star, and
$M\sub{targ}$ and $R\sub{targ}$ are its mass and
radius. $M\sub{field}$ is the mass of the star producing the tidal
field. Moments of inertia ${\rm MoI}\equiv I/MR^2$ have been
calculated from stellar evolution models \citep{Claret90}. ZAMS values
values of ${\rm MoI}=0.08$ for solar-mass stars, ${\rm MoI}=0.1$ for a
0.8 $M_\odot$ star and ${\rm MoI}=0.14$ for 0.6 $M_\odot$ stars have
been used to interpolate the value of this parameter for other masses.
For less massive stars we use values close to 0.23, which is the limit
for less centrally concentrated substellar objects \citep{Leconte11}.

Since we are interested in low mass binaries, i.e. $M_\star<1.5
\Msun$, for which convection happens in the whole star or in the outer
envelope \citep{Baraffe97}, we assume that turbulent convection is the
dominant tidal dissipation mechanism.  We compute the viscous
dissipation time $t\sub{diss}$ directly from convection overturn times
following Eq. (2.8) in \citet{Zahn08}.

For non-negligible eccentricities, tidal breaking drives stars to a
pseudo-synchronous final state where rotational angular velocity is
not exactly the average orbital angular velocity.  Final rotational
velocity in eccentric binaries is obtained when the term in the square
brackets in Eq. (\ref{eq:tsync}) becomes zero.

\begin{equation}
n\sub{sync}\equiv\frac{\Omega\sub{sync}}{\omega}=
\frac{1+\frac{15}{2}e^{2}+\frac{45}{8}e^{4}+\frac{5}{16}e^{6}}
     {(1-e^{2})^{\frac{3}{2}}(1+3e^{2}+\frac{3}{8}e^{4})}
\end{equation}

For eccentricities in the range 0-0.5, $1<n\sub{sync}<2.8$. 
Note that this equation is also Eq. (42) of \citet{Hut81}, where we use
average angular velocity rather than instantaneous periastron velocity.

\subsection{Planetary magnetospheres}
\label{subsec:PM}

For magnetic protection, we use the models of \citet{Zuluaga13}. 
Magnetosphere sizes, quantified by standoff distance $R_S$, 
are scaled with stellar wind dynamical pressure $P\sub{sw}=m n\sub{sw}
v\sub{sw}^2$ and the planetary dipole moment ${\cal M}$ according
to:

\beq{eq:RS}
\frac{R_S}{R_\oplus} = 9.75 \left(\frac{\cal
  M}{{\cal M}_\oplus}\right)^{1/3}
\left(\frac{P\sub{sw}}{P\sub{sw\odot}}\right)^{-1/6}
\eeq

where ${\cal M}_\oplus = 7.768\times 10^{22}$ A m$^2$ and
$P\sub{sw\odot}=2.24\times 10^{-9}$ Pa are the present Earth dipole
moment and the average dynamical pressure of the solar wind at
Earth.  We note that during the first
two Gyr, our thermal evolution models predict a dipole moment for
Earth analogues of around 0.6 ${\cal M}_\oplus$ (see Figure 4 in
\citealt{Zuluaga13}).

Strong magnetic fields were probably required to create magnetosphere
cavities large enough to protect early bloated atmospheres. For a
magnetosphere comparable in size, or smaller than, that predicted for
early Venus ($R\sub{S,Venus}$=3.8 $R_p$), subject to similar levels of
XUV radiation, we assume magnetic protection is insufficient to
prevent water loss similar to Venus.  On the other hand, if
magnetospheres are larger than that of early Earth
$R\sub{S,Earth}$=4.5 $R_p$ the planet is magnetically protected.

Earth and Venus limits are drawn in the contour plots for $R_S$ in
Figures \ref{fig:FXUV-Rs-1} and \ref{fig:FXUV-Rs-2}.  We stress that
while $R_S$ depends on the dipole moment, an intrinsic planetary
property, it also depends on the stellar wind pressure and velocity.
This is the reason why we can display $R_S$ in the $e-P_{bin}$ plane
for Earth-like magnetic properties.

\begin{figure}
\epsscale{1.15}
\plotone{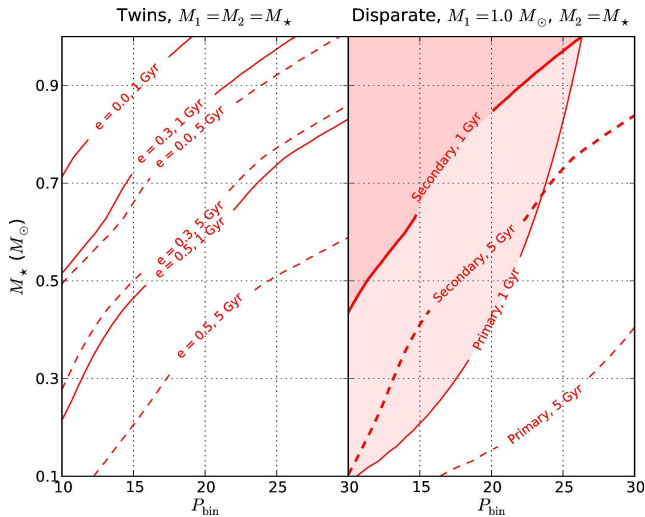}
\caption{Rotational synchronization times. The mass of twin stars
  (left) and the mass of the secondary, with a solar type primary,
  (right) is on the vertical axis.  Left: 1 Gyr (solid lines) and 5
  Gyr (dashed lines) synchronization times for eccentricities 0.1,
  0.3, and 0.5 are shown. Stars in binaries with high eccentricities
  become rotationally synchronized quicker than those with low
  eccentricities. Right: shaded regions show where one or both stars
  become tidally locked during the first Gyr.}
\label{fig:synctime}
\end{figure}

\section{Results and discussion}
\label{sec:results}

\begin{figure*}[t]
\epsscale{1.15}
\plottwo{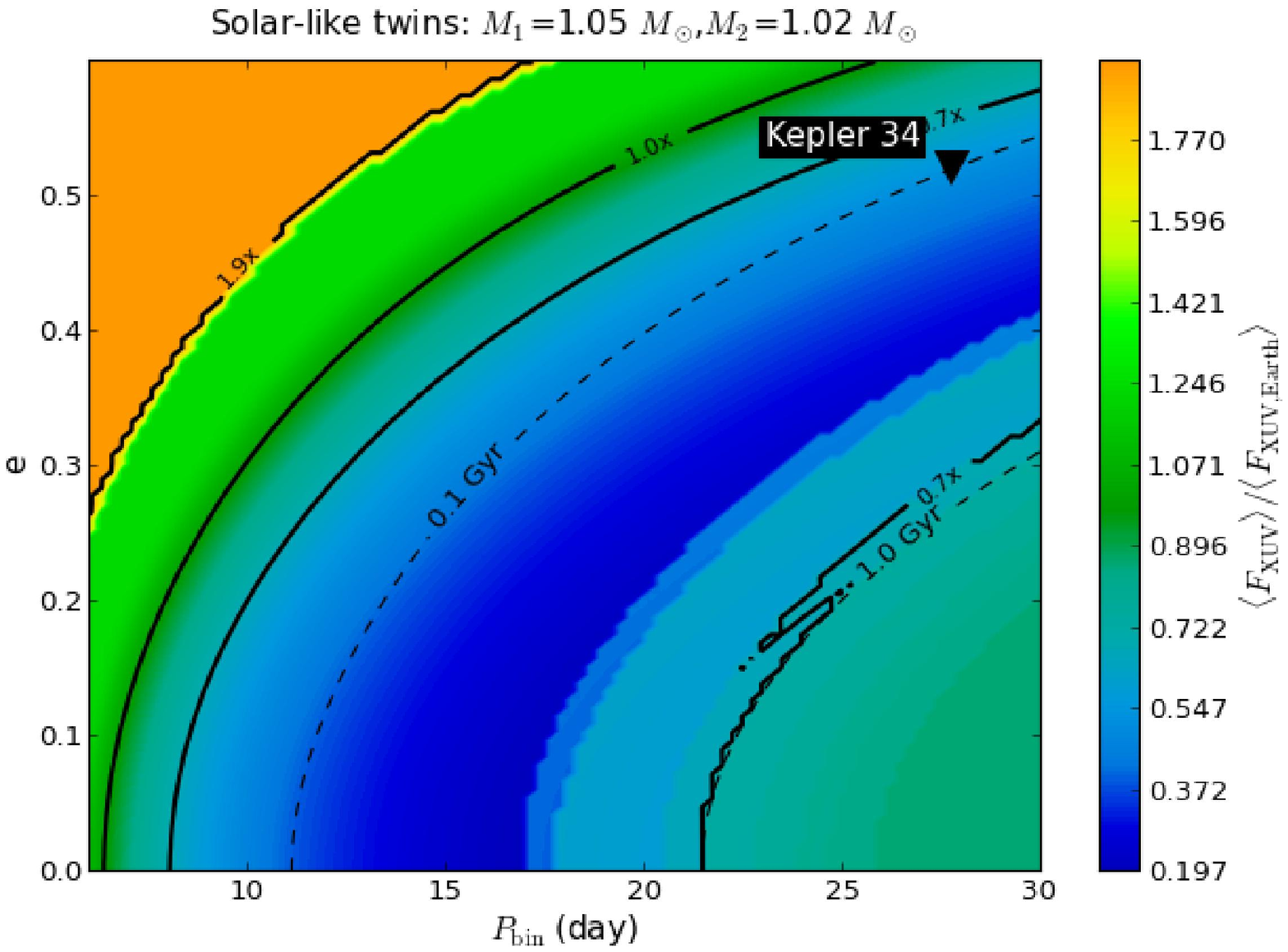}{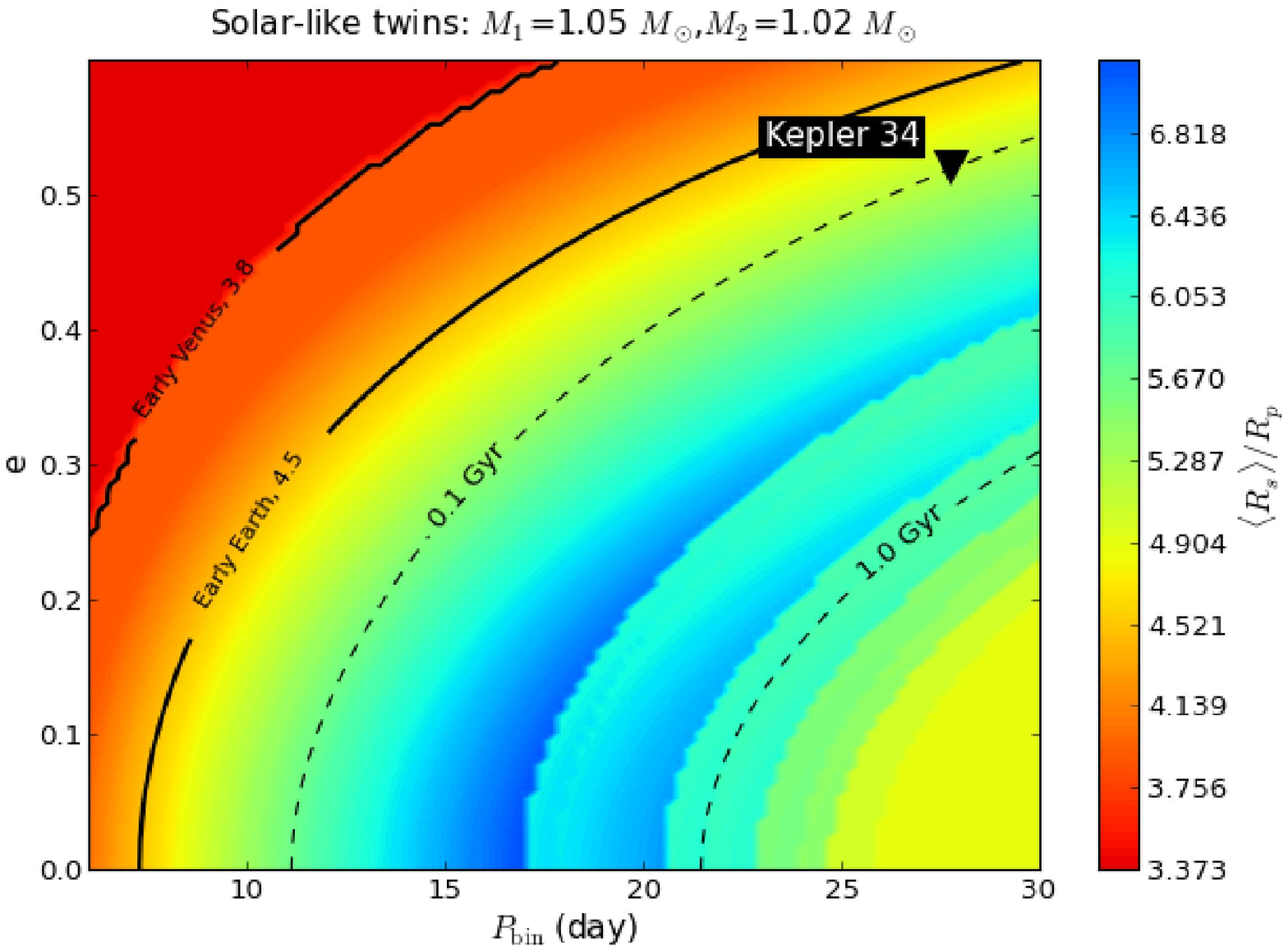}
\plottwo{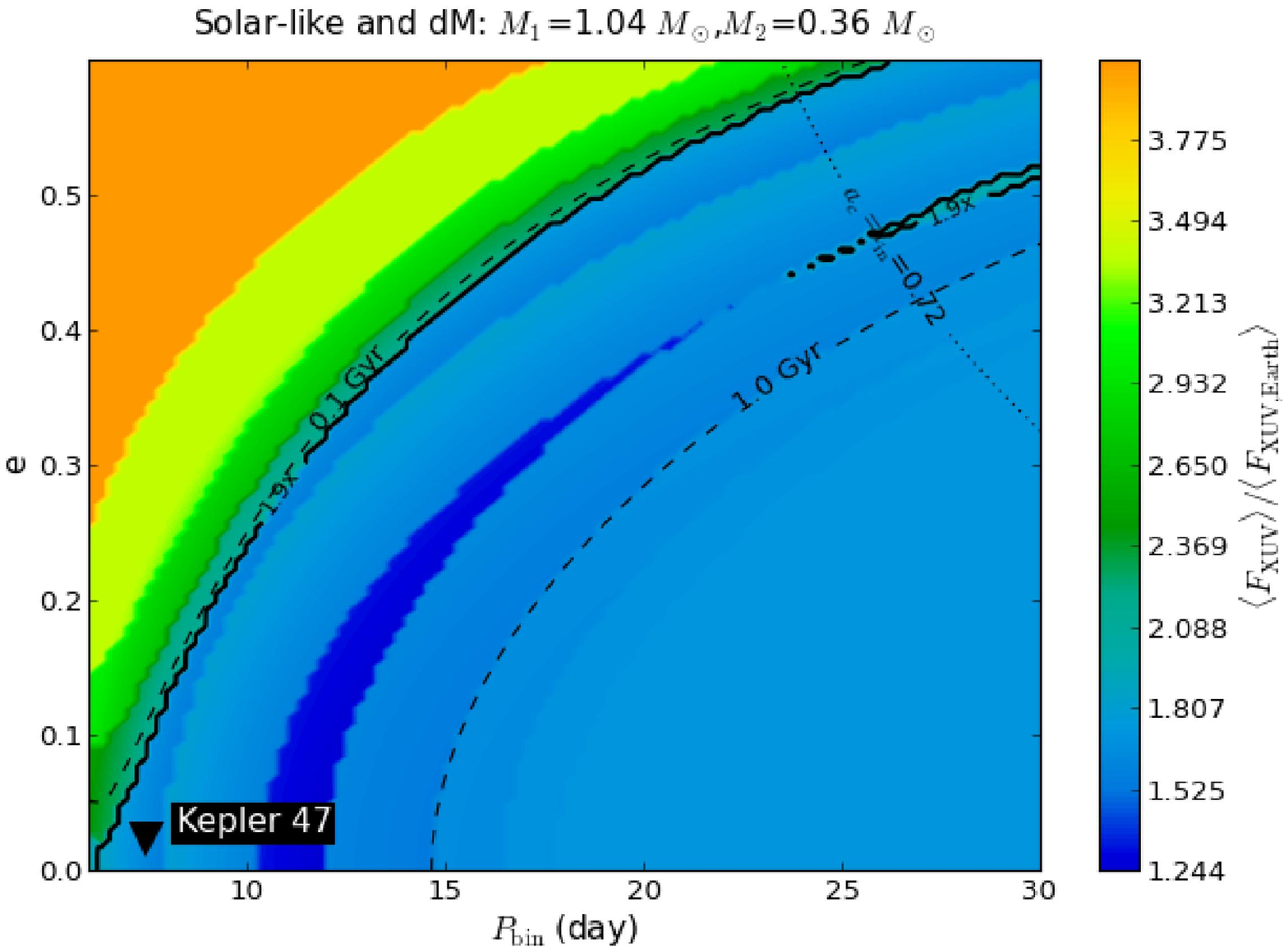}{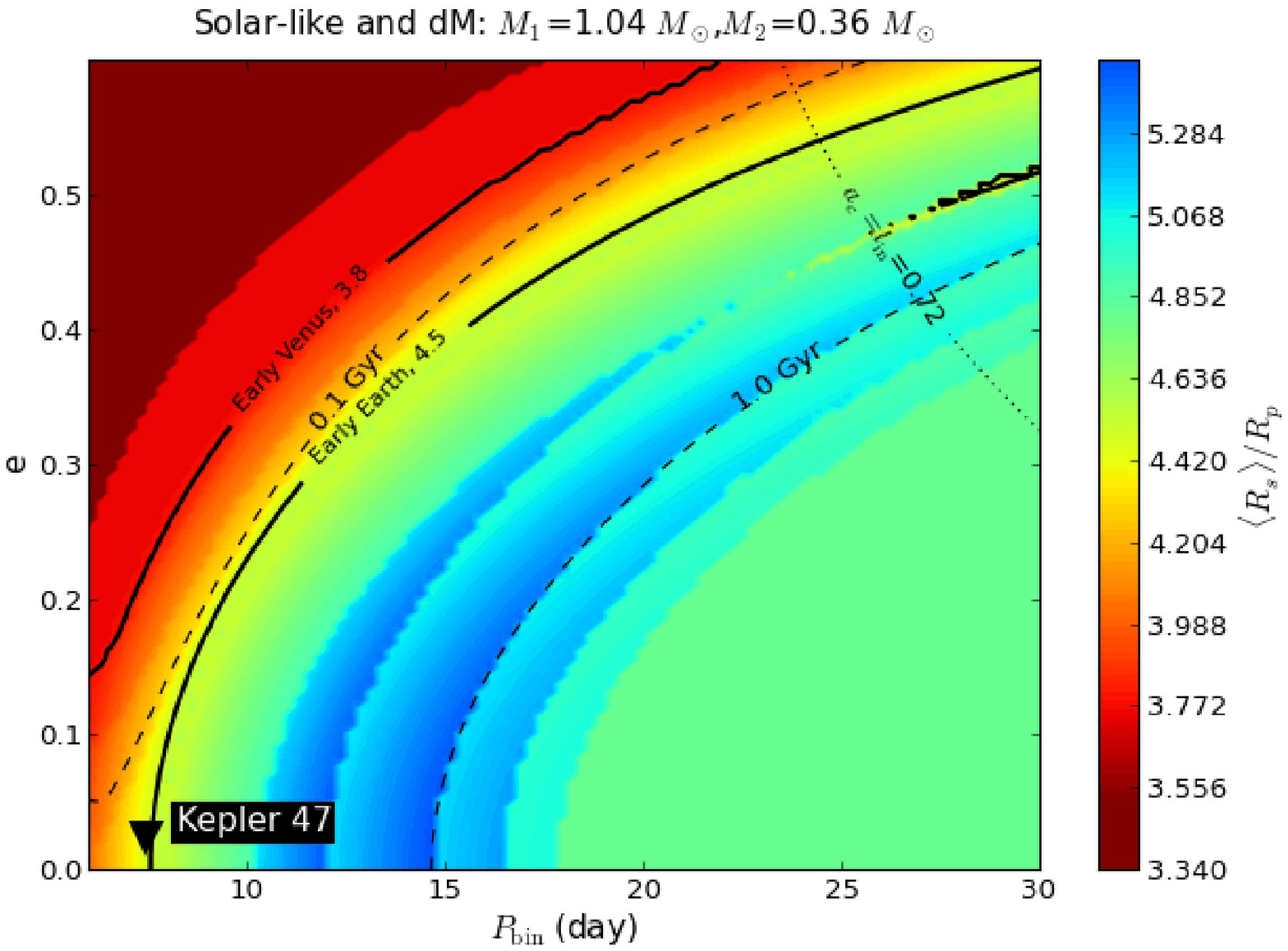}
\plottwo{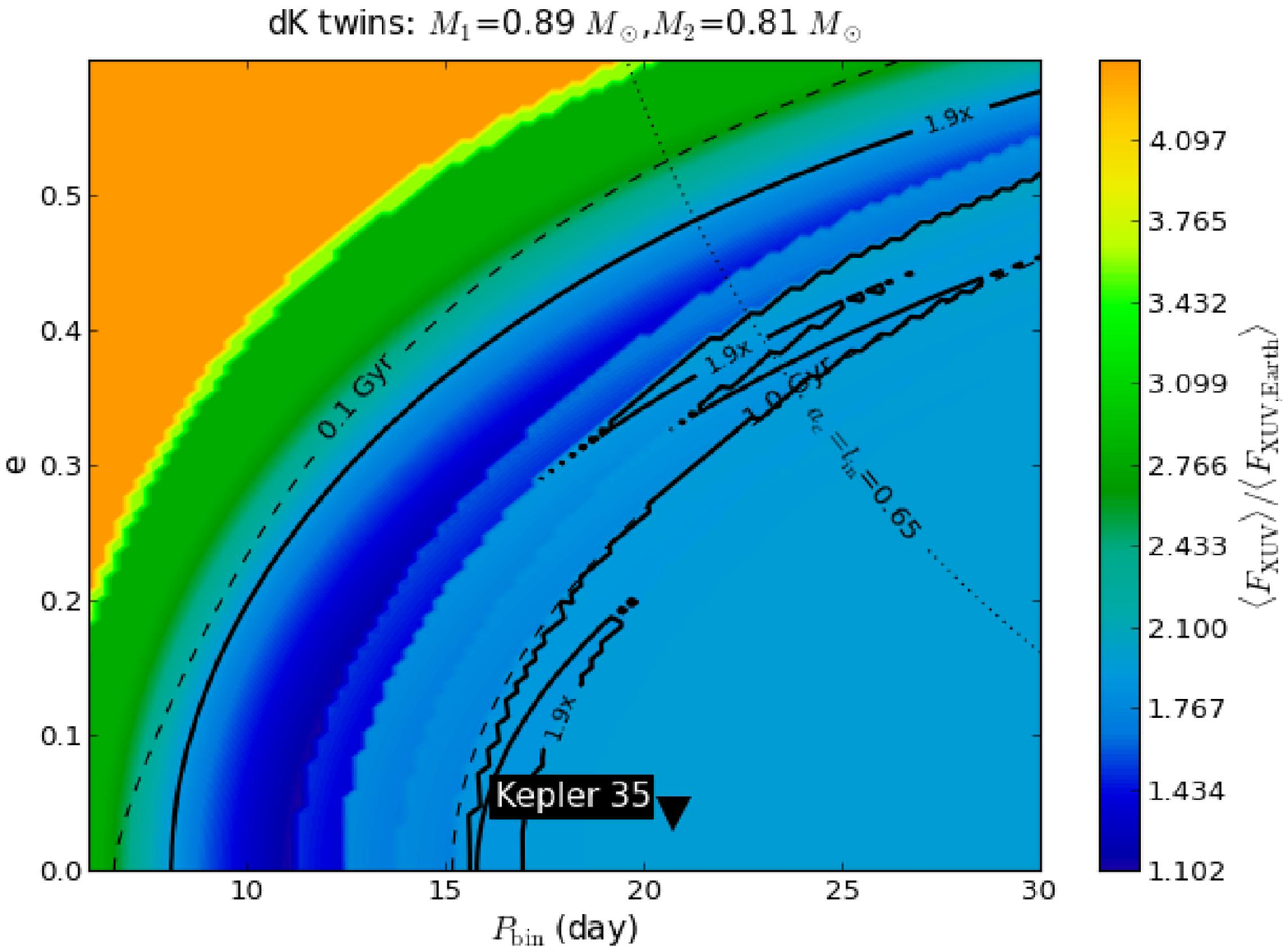}{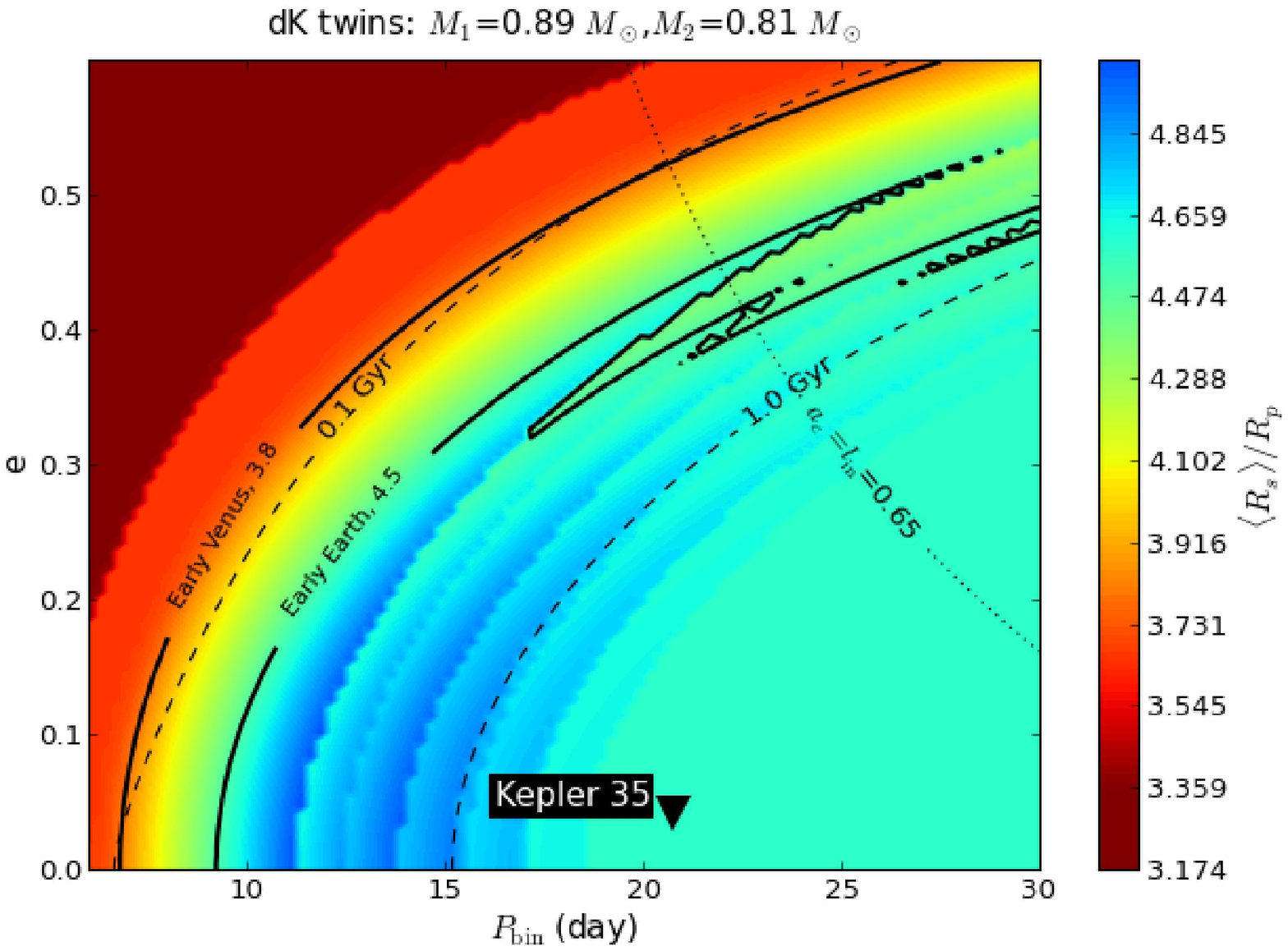}
\caption{Kepler binaries with enhanced habitability in the
  $e-P\sub{bin}$ plane. Binaries above and to the left of the 1Gyr
  lines have primaries that synchronize in less than 1 Gyr.  Those to
  the right and above and the dotted lines do not have stable orbits in
  the habitable zone.}
\label{fig:FXUV-Rs-1}    
\end{figure*}

\begin{figure*}[t]
\epsscale{1.15}
\plottwo{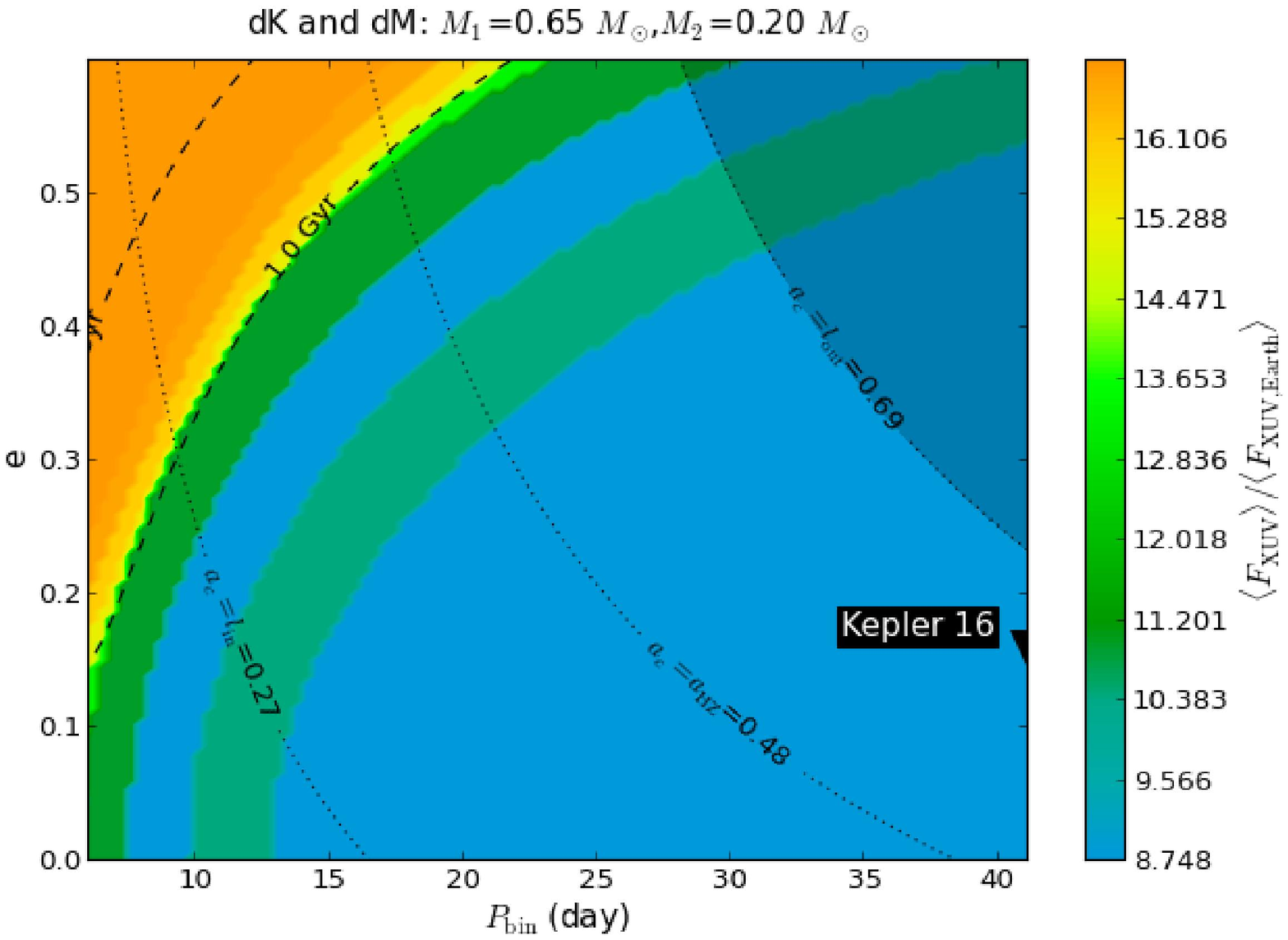}{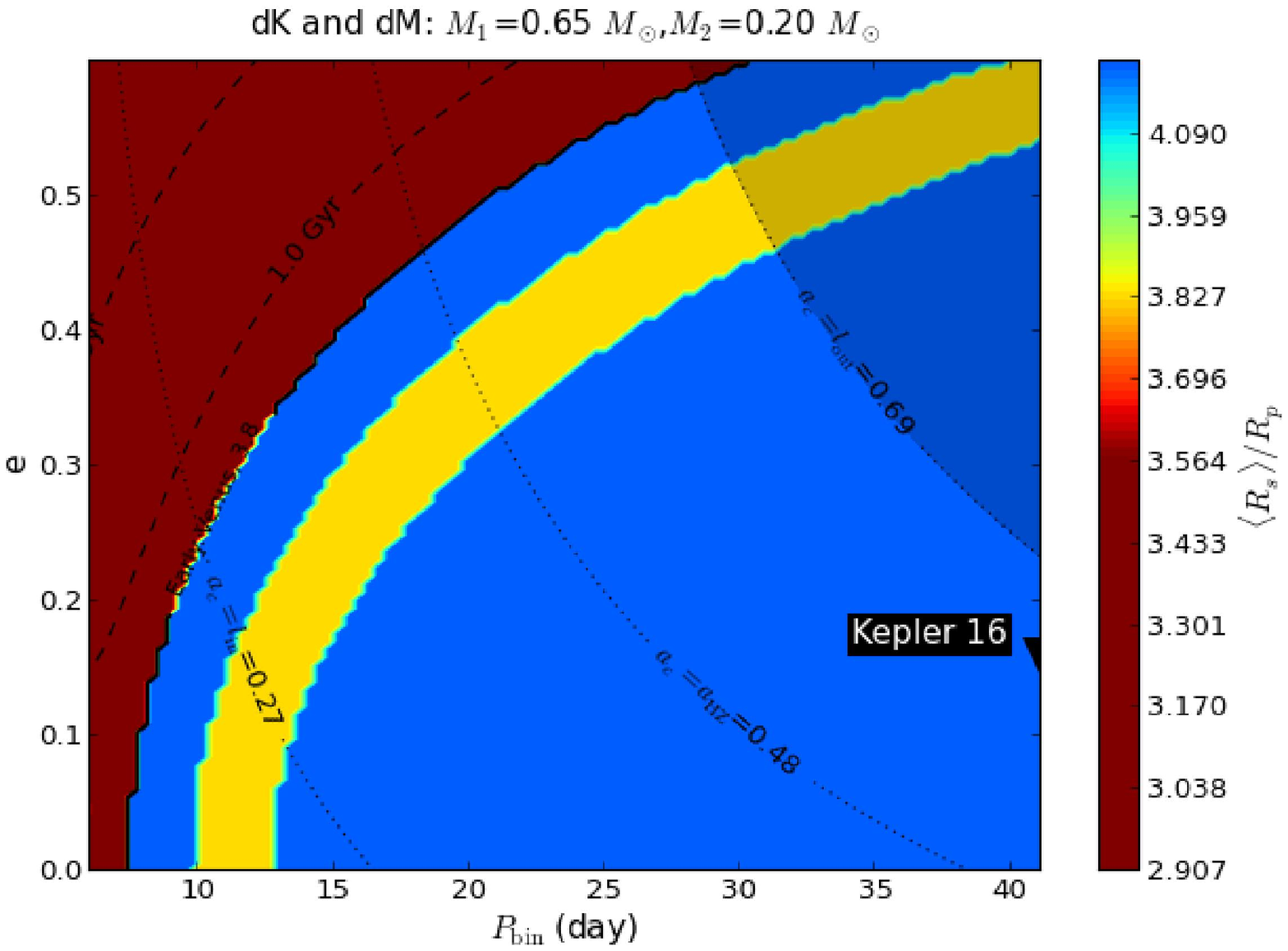}
\plottwo{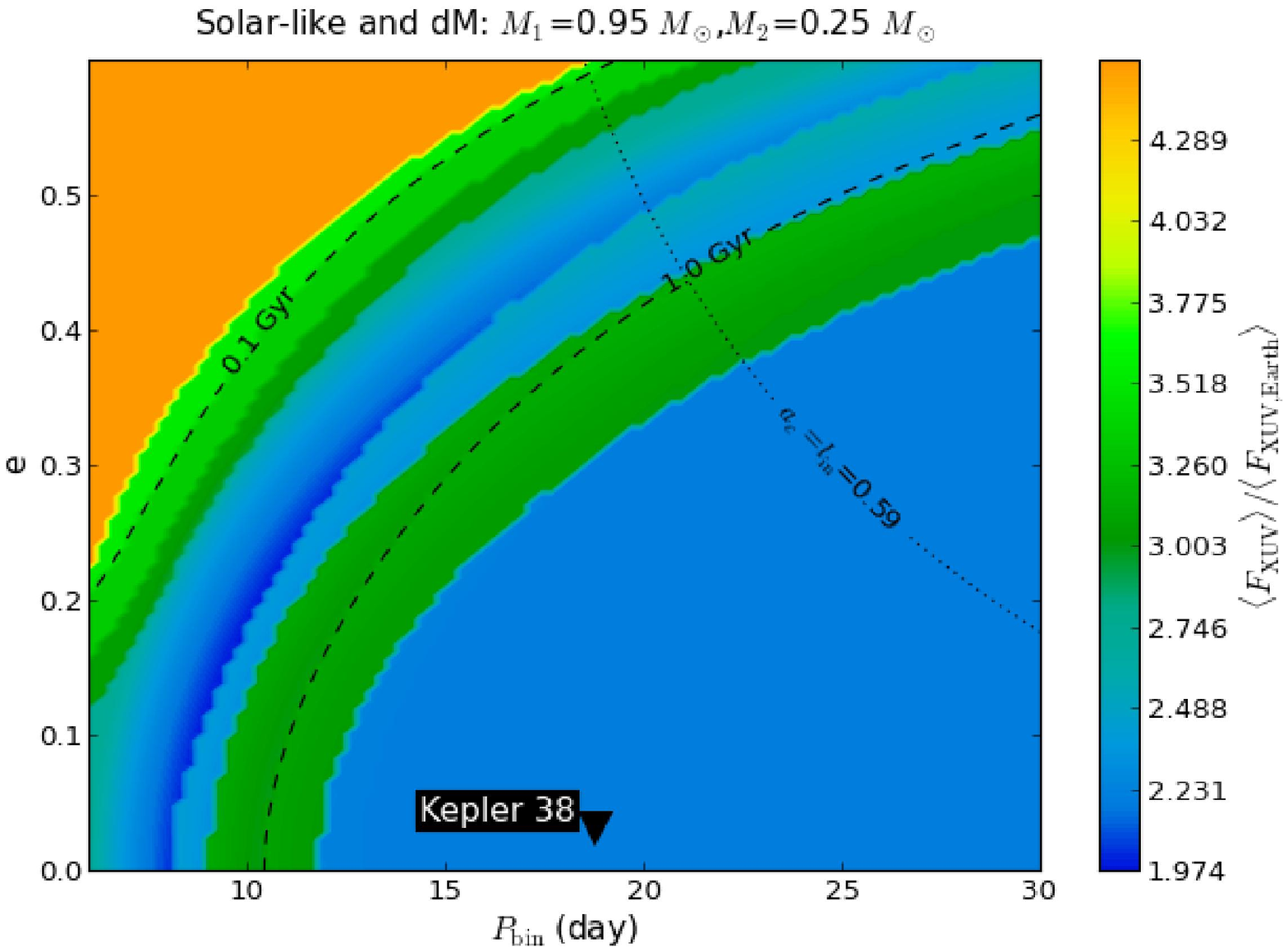}{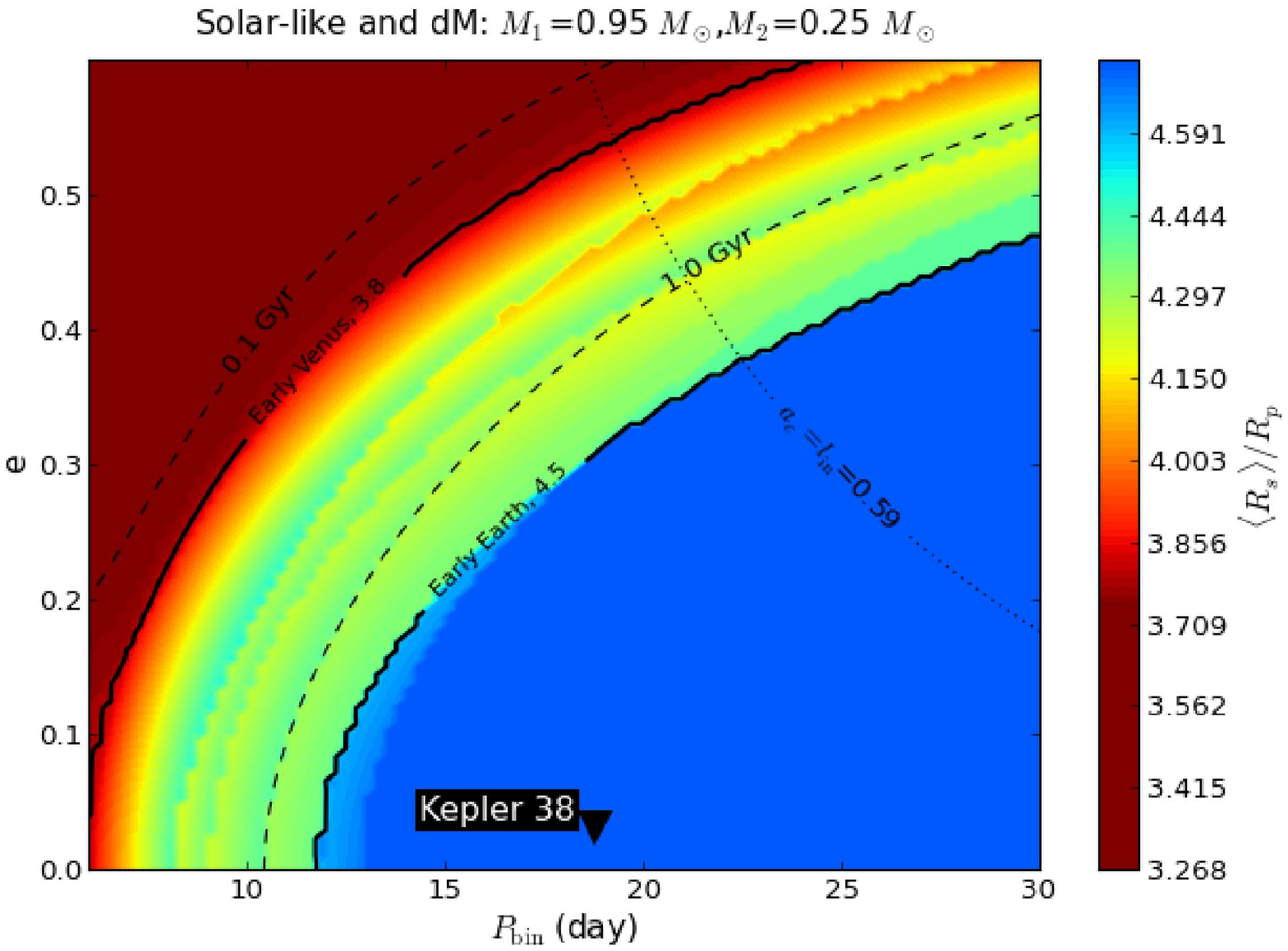}
\plottwo{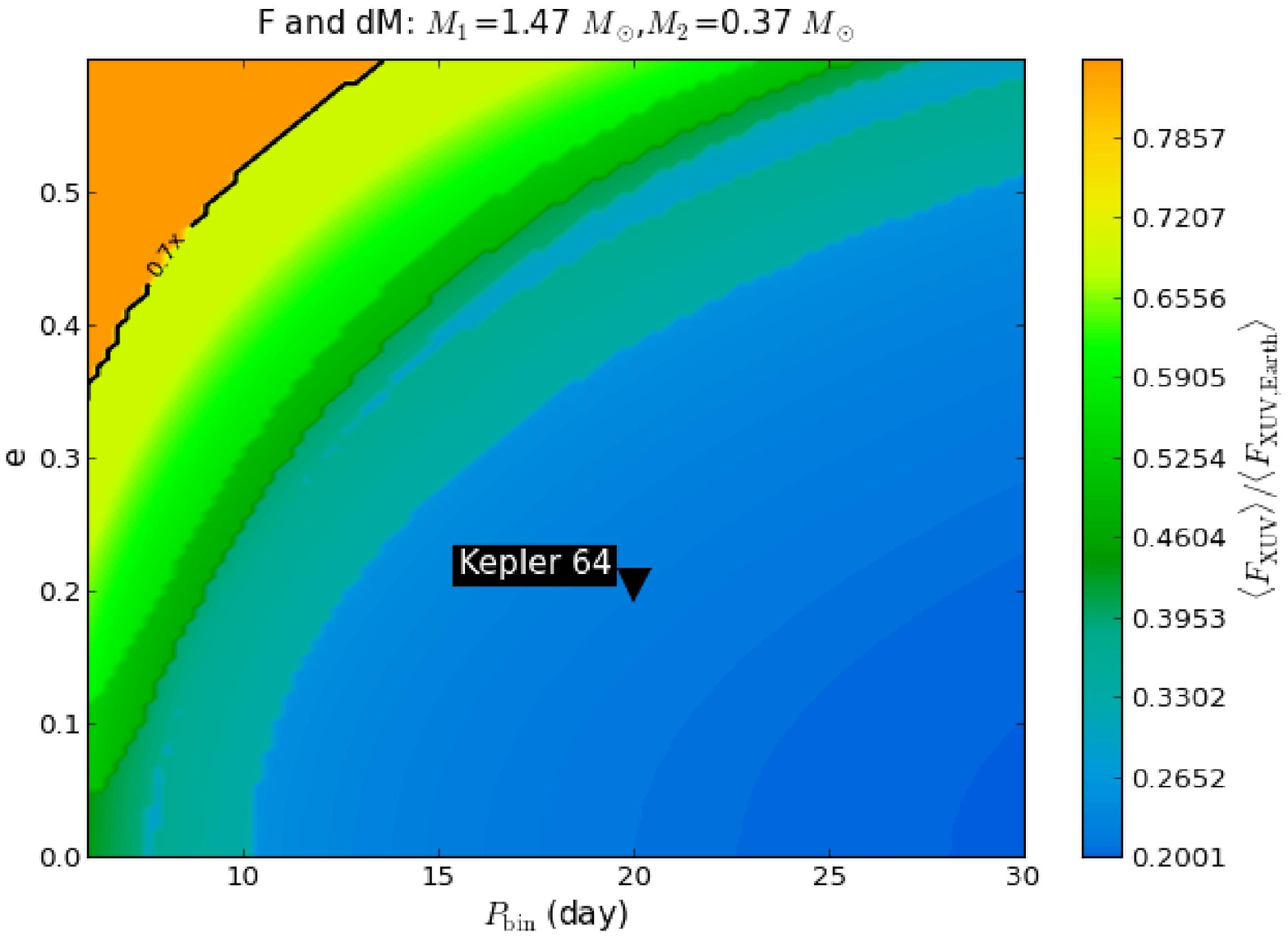}{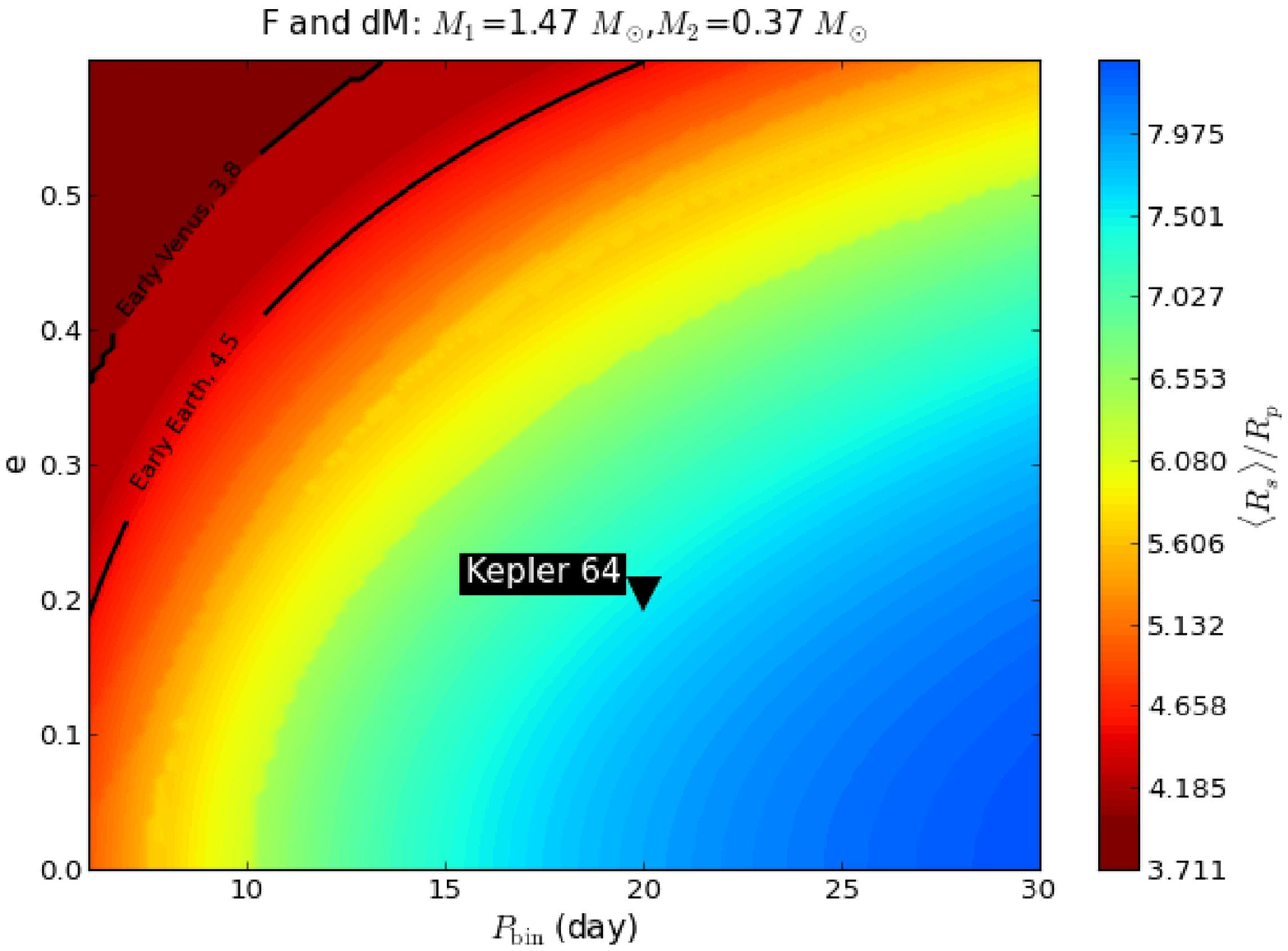}
\caption{Same as Figure 4, for Kepler 16, 38, and 64 (see text for details).}
\label{fig:FXUV-Rs-2}    
\end{figure*}

Relevant models are applied to study tidal synchronization effects on
circumbinary habitability.  Results for Kepler 34, Kepler 47 and
Kepler 35, showing enhanced habitability are given in Figure
\ref{fig:FXUV-Rs-1}.  Kepler 16, Kepler 38 and Kepler 64 results, are
given in Figure \ref{fig:FXUV-Rs-2}.  Kepler binaries (black
triangles) are plotted along with the level of stellar aggression
experienced by planets in the middle of the RHZ.  Dark blue areas in
the $e-P_{bin}$ plane correspond to binaries with reduced early XUV
flux and enhanced magnetic protection (large $R_S$).  We stress that
known Kepler binary system planets are not studied here, rather
hypothetical Earth-like planets in circumbinary habitable zones are
illustrated.

Solar mass twins, like Kepler 34 (black triangle in upper left panel of Figure
\ref{fig:FXUV-Rs-1}), for example, provide a RHZ that has $\sim$ 60\%
of the averaged XUV flux of early Earth $\langle
F\sub{XUV,Earth}\rangle$.  However, there is a spot, a binary
habitable niche, at $P\sub{bin}\sim 15$ days and $e=0$ for which a
mid-RHZ potentially habitable planet 
receives merely 20\% of the XUV flux of early Earth. Fluxes less than that
experienced by Venus exist for planets at mid-RHZ for all but the
shortest period and highest eccentricity solar twins (orange
region). Magnetospheric stand-off radii show a similar trend.  
Binaries that synchronize in less than 1 Gyr may provide
reduced stellar aggression.  However, this is true
only if the synchronization time is not too short (see the upper left corner
of the $e-P_{bin}$ plane) due to the forever young effect.
In fact, Earth-like conditions exist near the inner edge of the RHZ for solar-like
twins. A Venus-like planet with less magnetic protection than 
Earth could potentially maintain habitability in a system that could
also have an Earth-twin and even a water world farther out in
the RHZ. Hence multi-planet habitability is possible.

Lower mass twins or binaries with solar-like plus low mass companions,
Kepler 35 and Kepler 47 respectively, provide habitable conditions for
certain binary period and eccentricity combinations.  However, the
trend is towards less habitable conditions for lower mass binaries.
Kepler 35 consists of 0.89 $M_{\sun}$ and 0.81 $M_{\sun}$ stars in a nearly
circular orbit with a 21 day period. Its RHZ is exposed to about twice
the XUV flux as early Earth, but still is magnetically protected. The
same result applies in Kepler 47.  Binary habitable niches
for analogues of those systems are found around $P\sub{bin}\sim 12$
days and $e\sim 0.2$.

Results for Kepler 16, Kepler 38 and Kepler 64
binaries are shown in Figure \ref{fig:FXUV-Rs-2}.  These range from
the very inhospitable, Kepler 16 with two low mass stars, to the
planet friendly Kepler 64. A dwarf K and M pair (dK and dM), like
Kepler 16, have RHZs which are exposed to much higher XUV flux than
Earth. Kepler 38 ($M_{1} = 0.98 M_{\sun}$ and $M_{2} = 0.25 M_{\sun}$)
is a marginal case with Venus-like conditions at best. Single dM stars
are generally considered inhabitable based on their deadly XUV flux,
however, if paired with a solar companion, the dM star may synchronize
the primary, thereby providing a reduction of XUV flux. Kepler 64
($M_{1} = 1.53 M_{\sun}$ and $M_{2} = 0.38 M_{\sun}$) provides reduced
XUV flux and enhanced magnetic protection.  It is however,
limited by the relatively short lifetime of the primary star.

\bigskip

Our results show that if tidal-locking is capable of rotationally
synchronizing stars in binaries within the first Gyr, then the RHZs
may have reduced XUV and stellar wind flux. Planets with Earth-like
magnetic fields in the RHZ of these binaries will likely retain
atmospheric water. This effect is especially strong for solar-like (or
larger) primaries.  Planets that are less magnetically protected than
the Earth may survive desiccation even in the inner region of the
binary habitable zone.

Not only is it possible for Earth-like planets to exist in
circumbinary orbits for a wide range of binary parameters, but also
atmospheres experiencing less erosion than Earth are possible. This
has implications for both the number of habitable planets in the
Galaxy as well as the number of habitable planets per stellar system.
We suggest that the paradigm that binaries are not as suitable as
single stars for life, should be shifted to include a significant
number of potentially habitable circumbinary planets.

\acknowledgments

For additional material and updates please visit {\small
  \url{http://astronomia.udea.edu.co/binary-habitability}}.  We
appreciate comments by Ren\'e Heller and an anonymous referee. This
research is supported in part by NSF grant 0958783.  J.I. Zuluaga and
P.A. Cuartas-Restrepo are supported by CODI-UdeA.


\clearpage
\clearpage

\end{document}